\documentstyle[12pt]{article}

\newcommand{\mysection}[1]{\section{#1}\setcounter{equation}{0}}

\def\bea{\begin{eqnarray}} 
\def\eea{\end{eqnarray}}
\def\beann{\begin{eqnarray*}} 
\def\eeann{\end{eqnarray*}}
\def\beq{\begin{equation}} 
\def\eeq{\end{equation}}
  
\def\6{\partial } 
\def\7{\hat } 
\def\4{\tilde }

\def\gh{\mbox{gh}} 
\def\tot{\mbox{totdeg}}
\def\deg{\mbox{formdeg}}

\def\cA{{\cal A}}

\def\cC{{\cal C}}
\def\cD{{\cal D}}
\def\cE{{\cal E}}
\def\cF{{\cal F}}
\def\cL{{\cal L}}
\def\cR{{\cal R}} 
\def\cT{{\cal T}} 
\def\cU{{\cal U}} 
\def\cV{{\cal V}}
\def\cW{{\cal W}}

\begin{document}

\begin{flushright}
{\bf International Conference on}\\
{\bf Secondary calculus and cohomological Physics,}\\
{\bf Moscow, August 1997}
\end{flushright}

\begin{flushright}
UB-ECM-PF-97/36\\
hep-th/9711171\\
\end{flushright}

\begin{center}
{\Large Gauge Covariant Algebras and Local BRST Cohomology}
\end{center}

\begin{center}
{\large Friedemann Brandt}
\end{center}

\begin{center}
Departament ECM,
Facultat de F\'{\i}sica,
Universitat de Barcelona,\\
Diagonal 647,
E-08028 Barcelona, Spain
\end{center}

\begin{abstract}
The formulation of the local BRST cohomology 
on infinite jet bundles and its relation and reduction
to gauge covariant algebras are reviewed. As an illustration,
we compute the local BRST cohomology for geodesic motion
in (pseudo-) Riemannian manifolds and discuss briefly the result
(symmetries, constants of the motion, consistent deformations).
\end{abstract}

\mysection{Introduction}

The BRST formalism, invented by Becchi, Rouet and
Stora \cite{brs} for standard quantum gauge field theories,
can be established for general gauge theories. 
Its most useful and elegant formulation
uses ``fields'' and ``antifields'' \cite{bv} and goes also under 
the names ``BV'', ``field-antifield'' or just ``antifield'' formalism,
for reviews see e.g.\ \cite{henteit,report}. 

The outstanding feature of the formalism is to encode the equations of
motion, the gauge symmetries and their algebra in a single 
antiderivation which strictly squares to zero
on all fields and antifields. This antiderivation is 
called the BRST operator and denoted by $s$ here. Its existence 
allows us in particular to establish the local BRST cohomology. 
As pointed out and utilized already in \cite{brs},
this permits a cohomological analysis
of the anomaly problem in quantized gauge theories which initiated
an intense study of this aspect of the cohomology.
By now a number of further useful applications are known
in classical and quantum physics, as well as in the
theory of differential equations. 

One of these applications is the construction and classification of
consistent deformations of gauge invariant action
functionals and of their gauge symmetries \cite{bh,jds}.
Another one is the classification of rigid symmetries
and dynamical conservation laws as they correspond to cohomology
classes at negative ghost numbers \cite{bbh1}.
This includes variational symmetries \cite{olver} and the
characteristic cohomology of differential equations \cite{bryant}
(the BRST formalism can be established already
for differential equations which do not derive
from a Lagrangian, see e.g.\ \cite{banff}).  
Furthermore, the BRST operator can be extended so as to
incorporate not only the equations of motion, gauge symmetries etc., 
but the rigid symmetries too \cite{bhw}. This allows one for instance 
to determine analogously consistent deformations preserving a 
set of rigid symmetries, or deformations of
rigid symmetries themselves. We shall not consider this extension of the
BRST cohomology in the following, though the considerations can
be amended to treat it accordingly.

In the following we review and illustrate
an approach \cite{ten} to the local BRST cohomology which is based
on a few simple and universal concepts explained in more detail later:
\begin{enumerate}
\item
The formulation of the cohomological problem in the infinite
jet bundle of the fields and antifields.
\item
The mapping of this cohomological problem
to the cohomology of $\4s=s+d$ via the so-called descent equations.
\item
The construction of suitable local jet coordinates
simplifying the cohomological analysis considerably.
\end{enumerate}
A particularly interesting and useful
feature of this approach is to make contact
with differential geometric concepts, such as connections,
covariant derivatives and tensor calculus. 

The jet bundle formulation of the BRST cohomology 
and the descent equation technique
are reviewed in section \ref{jets}. 
Appropriate local jet coordinates and the relation to
differential geometric concepts are described in section \ref{tensors}. 
Finally the scheme is illustrated in section \ref{example}
by an explicit computation and discussion of the local BRST cohomology
for the geodesic motion of classical particles in 
Riemannian (or pseudo-Riemannian) manifolds.

\mysection{The BRST double complex in jet bundles}\label{jets}

In many applications one is interested in
the cohomology of the BRST operator in
the space of local functionals. Thereby a local functional is the
integral over a manifold ${\cal M}$ of a local function
of the fields and antifields, and two local functionals are 
identified if the Euler-Lagrange derivatives of their integrands
with respect to all fields and antifields agree (i.e.\ if their
integrands differ by a total derivative).
A local functional is called BRST invariant
when the BRST transformation of its integrand is a total derivative.
This cohomological problem can be defined accurately
in the space of differential forms on
the infinite jet bundle of the fields and antifields, which
is also the suitable arena to formulate and analyse the 
BRST cohomology for differential equations.
In this approach, one
views the fields and antifields as sections of
a fiber bundle $\pi: E\rightarrow {\cal M}$ and
prolongs $\pi$ to the associated infinite jet bundle
$J^\infty (E) \rightarrow {\cal M}$.
Standard local coordinates on $J^\infty (E)$ are local
coordinates $x^\mu$ of ${\cal M}$ supplemented by infinitely many 
coordinates corresponding to the partial derivatives of the 
fields $\Phi$ and antifields $\Phi^*$ 
(and turning into these partial derivatives on 
local sections of $E$). We denote these standard coordinates by
\beq 
\{Z^M\}\equiv \{x^\mu\, ,\, \6_{\mu_1\cdots\mu_k}\Phi^A\, ,\, 
\6_{\mu_1\cdots\mu_k}\Phi^*_\Delta:\ k=0,1,\dots\}.
\label{standard}\eeq
In many physically important cases fields and antifields correspond
one-to-one (are ``conjugate'') because the equations of motion
derive from an action functional via the variational principle.

In terms of the local jet coordinates (\ref{standard}),
$p$-forms on $J^\infty (E)$ are
\beq
\omega_p=\frac 1{p!}\, dx^{\mu_1}\dots dx^{\mu_p}
a_{\mu_1\dots\mu_p}(Z)
\label{pform}\eeq
where the differentials
are treated as anticommuting (Grassmann odd) objects, 
\[
dx^\mu dx^\nu = -dx^\nu dx^\mu.
\]
The cohomological problem is usually defined in a 
subspace of such forms, adapted to the problem under study
and referred to as the space
of {\em local} forms in the following. For instance,
one often requires that each $a_{\mu_1\dots\mu_p}(Z)$ in (\ref{pform})
is a function on some finite jet $J^k(E)$ 
where $k$ is arbitrary.
For definiteness we presuppose this definition of local forms 
in the following (except in section \ref{example}), but the
scheme is not restricted to it.

The BRST operator $s$ and the exterior derivative $d$
are antiderivations in the space of local forms.
They square to zero and anticommute,
\beq 
s^2=0,\quad s d+d s=0,\quad d^2=0.
\label{i1}\eeq
The exterior derivative is the standard one on $J^\infty(E)$,
\beq
d=dx^\mu \6_\mu\ ,
\label{d}\eeq
where $\6_\mu$ are total derivative operators
which read, in terms of the local coordinates (\ref{standard}),
\beq
\6_\mu=\frac{\6}{\6 x^\mu}
+\sum_{k\geq 0}\left[\6_{\mu\mu_1\cdots\mu_k}\Phi^A
\frac{\6}{\6\, \6_{\mu_1\cdots\mu_k}\Phi^A}+
\6_{\mu\mu_1\cdots\mu_k}\Phi^*_\Delta
\frac{\6}{\6\, \6_{\mu_1\cdots\mu_k}\Phi^*_\Delta}\right].
\label{partial}\eeq

Each local jet coordinate $Z^M$ has a particular ghost number,
assigned to it according to the standard rules of the BRST formalism
(in general, there are positive, vanishing and negative ghost numbers).
We can therefore consider the spaces 
$\Omega^{g,p}\equiv\Omega^{g,p}(J^\infty (E))$ 
of local $p$-forms with ghost number $g$.
The BRST operator carries ghost number 1 and form degree 0, 
i.e.\ it is a mapping
from $\Omega^{g,p}$ to $\Omega^{g+1,p}$. The exterior derivative $d$
is a mapping from $\Omega^{g,p}$ to $\Omega^{g,p+1}$.
We can therefore define the double complex $(\Omega^{*,*},s,d)$,
\beq
\begin{array}{lllllllllll}
& & & & \uparrow s & & \uparrow s & & & & \uparrow s
\\
 & & 0 & \longrightarrow & \Omega^{1,0} &
\stackrel{d}{\longrightarrow} & \Omega^{1,1} & 
\stackrel{d}{\longrightarrow} &
\cdots & \stackrel{d}{\longrightarrow} & \Omega^{1,n}
\\
& & & & \uparrow s & & \uparrow s & & & & \uparrow s
\\
0 & \longrightarrow & \mbox{\bf R} & \longrightarrow & \Omega^{0,0} &
\stackrel{d}{\longrightarrow} & \Omega^{0,1} & 
\stackrel{d}{\longrightarrow} &
\cdots & \stackrel{d}{\longrightarrow} & \Omega^{0,n}
\\
& & & & \uparrow s & & \uparrow s & & & & \uparrow s
\\
 & & 0 & \longrightarrow & \Omega^{-1,0} &
\stackrel{d}{\longrightarrow} & \Omega^{-1,1} & 
\stackrel{d}{\longrightarrow} &
\cdots & \stackrel{d}{\longrightarrow} & \Omega^{-1,n}
\\
& & & & \uparrow s & & \uparrow s & & & & \uparrow s
\end{array}
\label{2complex}
\eeq
where $n=dim(M)$. In general, the diagram is
infinite in the vertical directions.

We define the local BRST cohomology as 
the cohomology of $s$ modulo $d$ at form degree $n$ and denote
it by $H(s|d,\Omega^{*,n})$.
The corresponding cocycle condition and 
equivalence relation ($\cong$) are 
\bea
& s\omega_n+d\omega_{n-1}=0, &
\label{des1}\\
& \omega_n\cong\omega'_n\quad :\Leftrightarrow\quad 
\omega_n-\omega'_n=s\eta_n+d\eta_{n-1}\ . &
\label{des2}\eea
A useful tool to analyse this cohomological problem are the
so-called descent equations,
\beq
s\omega_p+d\omega_{p-1}=0,\quad p=0,\dots,n\quad
(\mbox{with}\quad \omega_{-1}\equiv 0),
\label{des3}\eeq
where $\omega_p$ may vanish for all $p$ smaller than some 
$p_0$. (\ref{des3}) follows from (\ref{des1}),
because the cohomology of $d$ is
trivial in $\Omega^{g,p}$, at least locally, except for
$(g,p)=(0,0)$ and $p=n$ \cite{various},
\beq
H(d,\Omega^{g,p}) \simeq \delta^0_g\, \delta^0_p\, \mbox{{\bf R}}\quad 
\mbox{for}\quad p<n.
\label{Hd}\eeq
Using this result which goes in physics
sometimes under name ``algebraic Poincar\'e lemma'',
the descent equations (\ref{des3}) are derived as follows.
Acting with $s$ on (\ref{des1}) and using (\ref{i1}),
one obtains $d(s\omega_{n-1})=0$, i.e.\ $s\omega_{n-1}$ is
a closed form. (\ref{Hd}) thus implies the existence of
a local form $\omega_{n-2}$ satisfying $s\omega_{n-1}+d\omega_{n-2}=0$.
One now acts with $s$ on the latter equation,
repeats the arguments, and derives ultimately (\ref{des3}) 
in this manner\footnote{(\ref{Hd}) alone would still permit that the
equation for $p=0$ in (\ref{des3}) can read $s\omega_0=constant$
in the case that $s\omega_0$ has ghost number 0. However,
a nonvanishing constant in $s\omega_0$ would signal the inconsistency 
of the equations of motion and can
thus be excluded without loss of generality. Indeed, denoting
the equations of motion by $L_i=0$, $s\omega_0=constant$ would imply 
$K^i L_i=constant$ for some differential operators $K^i$ which
is clearly inconsistent unless the constant vanishes.}.

The descent equations (\ref{des3}) can be cast in the compact form
\beq
\4s\, \4\omega=0
\label{des4}\eeq
where $\4s$ is simply the sum of $s$ and $d$,
\beq
\4s=s+d,
\label{s+d}\eeq
and $\4\omega$ is the sum of the local $p$-forms occurring in the
descent equations (\ref{des3}),
\beq
\4\omega=\sum_{p=0}^n \omega_p\ .
\label{omega}\eeq
Furthermore (\ref{Hd}) implies that  
the equivalence relation (\ref{des2}) translates into
\beq
\4\omega\cong\4\omega'\quad :\Leftrightarrow\quad
\4\omega-\4\omega'=\4s\, \4\eta+constant
\label{des5}\eeq
where $\4\eta=\sum\eta_p$.
Indeed, assume that $\omega_n$ solves (\ref{des1}) trivially, i.e.
$\omega_n=s\eta_n+d\eta_{n-1}$. Then (\ref{des1}) implies
$d(\omega_{n-1}-s\eta_{n-1})=0$ because of (\ref{i1}). Using
(\ref{Hd}), we conclude $\omega_{n-1}=s\eta_{n-1}+d\eta_{n-2}$
for some $\eta_{n-2}$. Repeating the arguments one derives
$\omega_p=s\eta_p+d\eta_{p-1}$ for all $p>0$ and 
$\omega_0=s\eta_0+constant$ which altogether give 
$\4\omega=\4s\4\eta+constant$. Conversely $\4\omega=\4s\4\eta+constant$
implies of course $\omega_n=s\eta_n+d\eta_{n-1}$. Hence,
(\ref{des1}) and (\ref{des2}) are indeed equivalent to
(\ref{des4}) and (\ref{des5}), at least locally.

Note that $\4s$ is an antiderivation
which squares to zero by (\ref{i1}),
\beq \4s^2=0.
\label{4s2}\eeq
Notice also that the descent descent equations (\ref{des3})
concern diagonal directions in
the double complex (\ref{2complex}) for which the sum of the
ghost number ($\gh$) and the form degree ($\deg$) is constant. 
This sum is
of course the natural degree when working with $\4s$ and
is called ``total degree'' ($\tot$) in the following,
\beq
\tot=\gh+\deg .
\label{tot}\eeq
A sum of local forms as in (\ref{omega}) will be called
local ``total form'' in the following. We conclude that
the local BRST cohomology at ghost number $g$ is (locally) 
isomorphic to the cohomology of $\4s$ in the space of 
local total forms at total degree $(g+n)$,
\beq
H(s|d,\Omega^{g,n}) \simeq H(\4s,\4\Omega^{g+n}),\quad
\4\Omega^G=\bigoplus_{p=0}^n\Omega^{G-p,p}.
\label{iso1}\eeq

{\sc Remarks.} 
a) (\ref{Hd}) can fail to be globally valid.
If such global considerations are relevant, one must refine the 
analysis of the descent equations, cf.\,\cite{bbh3} for an example.

b) $H(\4s,\4\Omega^*)$ is isomorphic to an
antifield independent ``weak cohomology'' \cite{ten}.

c) In general, (\ref{iso1}) does not extend to 
$H(s|d,\Omega^{g,p})$ for $p<n$. Namely, locally
$s\omega_p+d\omega_{p-1}=0$ implies descent equations
of the form $s\omega_q+d\omega_{q-1}=0$, $q=0,\dots, p$,
but it does {\em not} guarantee the existence of
a local $(p+1)$-form $\omega_{p+1}$ satisfying
$s\omega_{p+1}+d\omega_p=0$. The reason is the following. 
Acting with $d$ on $s\omega_p+d\omega_{p-1}=0$,
one concludes $s(d\omega_p)=0$, i.e. $d\omega_p$
is an $s$-closed $(p+1)$-form. However,
$H(s,\Omega^{*,p+1})$ can be nontrivial, i.e.
$d\omega_p$ is not necessarily $s$-exact in the
space of local $(p+1)$-forms. 
In particular, $H(s|d,\Omega^{g,p})$ and
$H(s|d,\Omega^{g+1,p-1})$ are not isomorphic in general, not even
locally.

\mysection{Suitable jet coordinates and
relation to differential geometric concepts}\label{tensors}

In order to compute the cohomology of $\4s$ in practice,
it is very useful, if not indispensable, to
switch from the standard jet coordinates (\ref{standard}),
supplemented with the differentials\footnote{When working
in the space of local total forms, it is natural
and convenient to supplement (\ref{standard}) with the
differentials $dx^\mu$. This supplement is assumed in the following.},
to new ones which are better adapted to
the cohomological analysis.

Particularly useful
are local jet coordinates $\{\cU^\ell,\cV^\ell,\cW^i\}$
satisfying
\beq
\4s\, \cU^\ell=\cV^\ell,\quad
\4s\, \cW^i=\cR^i(\cW)
\label{e1}
\eeq
where the $\cR^i(\cW)$ are functions of the $\cW$'s only.
Such a change of local jet coordinates
is required to be compatible with the locality requirement
imposed on the particular cohomological problem,
i.e.\ the $\cU$'s, $\cV$'s and $\cW$'s 
themselves have to be local total forms such that
$\4\Omega^*$ factorizes locally
into the $\4s$-invariant subspaces 
$\4\Omega^*_{\cU,\cV}$ and $\4\Omega^*_{\cW}$
of local total forms depending only on the $\cU^\ell,\cV^\ell$
and on the $\cW^i$ respectively,
\[
\4\Omega^*=\4\Omega^*_{\cU,\cV} \times \4\Omega^*_{\cW}\ ,\quad
\4s\, \4\Omega^*_{\cU,\cV}\subseteq \4\Omega^*_{\cU,\cV}\ ,\quad
\4s\, \4\Omega^*_{\cW}\subseteq \4\Omega^*_{\cW}\ .
\]
Then, by K\"unneth's formula, the cohomology of $\4s$ on
$\4\Omega^*$ factorizes accordingly
\beq
H(\4s,\4\Omega^G)=\bigoplus_{G'}
H(\4s,\4\Omega^{G'}_{\cU,\cV}) \times H(\4s,\4\Omega^{G-G'}_{\cW}).
\label{kunneth}\eeq
Moreover, thanks to the first Eq.\ (\ref{e1}),
$H(\4s,\4\Omega^*_{\cU,\cV})$ is contractible,
at least locally. Hence, locally
the $\cU$'s and $\cV$'s drop out of the cohomology and we are
left with the cohomology of $\4s$ on local total forms constructed
only of the $\cW$'s,
\beq
H(\4s,\4\Omega^G_{\cU,\cV}) \simeq \delta_G^0\, \mbox{{\bf R}}
\quad\Rightarrow\quad
H(\4s,\4\Omega^G) \simeq H(\4s,\4\Omega^{G}_{\cW}).
\label{e2}\eeq
Therefore the $\cU$'s and $\cV$'s are called ``trivial pairs''.
Clearly the aim is to construct a set of new local jet coordinates
with as many trivial pairs as possible. Thereby the challenge is
usually not the finding of the $\cU$'s and $\cV$'s but the
construction of corresponding complementary $\cW$'s.
The latter are interpreted as generalized connections, tensor fields
and covariantized antifields. We shall try to motivate this terminology 
in the following by general arguments and
an example in the next section. However,
to fully understand and appreciate it, an inspection of
further examples is helpful, see e.g.\ \cite{ten} and Refs.\ therein.

For notational simplicity we will assume that the gauge transformations
are irreducible, i.e.\ that ghosts for ghosts are not needed.
Then all local jet coordinates (\ref{standard}) have
ghost numbers $\leq 1$. Consequently, each
$\cW^i$ can be assumed to have a definite total degree $\leq 1$, and
we can decompose the set of $\cW$'s into subsets with
definite total degree as follows:
\bea
\{\4\cC^N\}  &=& \{\cW^i:\ \tot(\cW^i)=1\} ,
\label{4C}\\
\{\4\cT^r\} &=& \{\cW^i:\ \tot(\cW^i)=0\} ,
\label{4T}\\
\{\4\cT_\delta^*\} &=& \{\cW^i:\ \tot(\cW^i)=-1\} ,\ \dots
\label{4*}\eea
For reasons explained below, we call
the $\4\cC$'s ``generalized connections'' and the
$\4\cT$'s ``generalized tensor fields''.
Because of (\ref{tot}) one has
\beq
\4\cC^N = \7C^N+dx^\mu \cA_\mu^N+\dots\quad , \quad
\4\cT^r = \cT^r+\dots 
\label{decCT}
\eeq
where dots denote antifield dependent terms, and
$\7C^N$, $\cA_\mu^N$ and $\cT^r$ depend only on the fields
and their derivatives and, possibly, explicitly
on the coordinates of the base manifold (but neither on antifields,
nor on differentials).
The $\7C^N$ have ghost number 1,
whereas the $\cA_\mu^N$ and $\cT^r$ have ghost number 0.
The antifield dependent terms in
$\4\cC^N$ and $\4\cT^r$ involve necessarily ghosts or
differentials.
Of course, the $\cW$'s with negative total degree
cannot have antifield independent parts. They are
``covariantized antifields''.

In order to explain the relation to 
differential geometric concepts, we spell
out more explicitly the second Eq.\ (\ref{e1}), using
(\ref{4C})--(\ref{4*}).
As $\4s$ has total degree 1, one gets
\bea
\4s\, \4\cC^M &=& \frac 12\, (-)^{\varepsilon_N+1}\4\cC^N \4\cC^P
                {\cF_{PN}}^M(\4\cT)
                + \frac 12\, (-)^{\varepsilon_P+1}\4\cT_\delta^*\,
                \4\cC^N \4\cC^P \4\cC^Q
                {\cE_{QPN}}^{\delta M}(\4\cT)
                + O(4),
\label{e3}\\
\4s\, \4\cT^r &=& \4\cC^M {\cR_M}^r(\4\cT)
                  + \frac 12\, (-)^{\varepsilon_M+1}
                  \4\cT_\delta^*\, \4\cC^M \4\cC^N
                  {\cE_{NM}}^{\delta r}(\4\cT)
                  + O(3),
\label{e4}\\
\4s\, \4\cT_\delta^* &=& \cL_\delta (\4\cT) + O(1),\quad\dots
\label{e5}
\eea
where $O(k)$ denotes terms of order $\geq k$ in the $\4\cC$'s,
and $(\varepsilon_M+1)$ is the Grassmann parity of $\4\cC^M$.
Due to $\4s^2=0$ the functions of the $\4\cT$'s appearing
in (\ref{e3})--(\ref{e5}) are related. In particular,
$\4s^2\4\cT^r=0$ and $\4s^2\4\cC^M=0$ yield at lowest degree
in the $\4\cC$'s respectively
\bea
&
\4\nabla_{[M}{\cR_{N]}}^r(\4\cT)
+{\cF_{MN}}^P(\4\cT) {\cR_P}^r(\4\cT)
+\cL_\delta(\4\cT) {\cE_{MN}}^{\delta r}(\4\cT)=0,
&
\label{e6}\\
&
\4\nabla_{[M}{\cF_{NP]}}^R(\4\cT)
-(-)^{\varepsilon_P\varepsilon_Q}{\cF_{[MN}}^Q(\4\cT){\cF_{P]Q}}^R(\4\cT)
+\cL_\delta(\4\cT) {\cE_{MNP}}^{\delta R}(\4\cT) = 0
&
\label{e7}\eea
where $[\cdots]$
denotes graded antisymmetrization, and we used the definition
\beq
\4\nabla_M:={\cR_M}^r(\4\cT)\, \frac{\6}{\6\4\cT^r}\ .
\label{e8}\eeq
It is instructive to look at the antifield independent
parts of (\ref{e3})--(\ref{e7}), using (\ref{decCT}).
{}From (\ref{e5}) one infers that $\cL_\delta(\cT)$
vanishes weakly (``on-shell''), i.e.\ it is ``a combination of the
left hand sides of the equations of motion'',
\beq
\cL_\delta(\cT) \approx 0.
\label{e9}\eeq
This follows because (\ref{e5}) implies that
the antifield independent part of $\cL_\delta(\4\cT)$
is in the image of the so-called
field theoretical Koszul-Tate differential 
(see e.g.\ \cite{henfisch,henteit}) contained in the BRST operator.
(\ref{e6}) and (\ref{e7}) yield respectively
\bea
&
\mbox{{\bf [}} \, \nabla_M\, ,\, \nabla_N\, \mbox{{\bf ]}}
\approx -{\cF_{MN}}^P(\cT) \nabla_P\ ,
&
\label{e10}\\
&
\nabla_{[M}{\cF_{NP]}}^R(\cT)
-(-)^{\varepsilon_P\varepsilon_Q}
{\cF_{[MN}}^Q(\cT){\cF_{P]Q}}^R(\cT) \approx 0
&
\label{e11}
\eea
where {\bf [}\ ,\ {\bf ]} denotes the graded commutator, and
\beq
\nabla_M:={\cR_M}^r(\cT)\, \frac{\6}{\6\cT^r}\ .
\label{e12}\eeq
According to (\ref{e10}) the (graded) commutator algebra
of the $\nabla_M$ closes in the weak sense (on-shell).
(\ref{e11}) is implied by this algebra (Jacobi identity).

The antifield independent part of (\ref{e4}) splits into
two parts with ghost numbers 1 and 0,
\bea
\gamma \cT^r & \approx & \7C^N \nabla_N \cT^r,
\label{e13}\\
\6_\mu \cT^r & \approx & \cA_\mu^N \nabla_N \cT^r,
\label{e14}
\eea
where $\gamma \cT^r$ is the antifield independent
part of $s \cT^r$. Here we used again that
antifield independent pieces in $s\4\cT$ vanish weakly
whenever they are in the image of the Koszul-Tate differential.
Analogously the antifield independent part of
(\ref{e3}) yields
\bea
& \gamma \7C^M \approx \frac 12(-)^{\varepsilon_N+1}\7C^N\7C^P
{\cF_{PN}}^M(\cT), &
\label{cov14}\\
& \gamma \cA^M_\mu \approx
\6_\mu\7C^M-\7C^N\cA^P_\mu {\cF_{PN}}^M(\cT), &
\label{cov15}\\
& \6_\mu \cA^M_\nu-\6_\nu \cA^M_\mu
+\cA^N_\mu \cA^P_\nu {\cF_{PN}}^M(\cT)  \approx 0.&
\label{cov16}\eea

To interpret the above equations,
we recall that $\gamma\cT^r$ equals the infinitesimal
gauge transformation of $\cT^r$ with gauge parameters replaced
by the ghost fields. (\ref{e13}) thus means that the gauge
transformations of the $\cT$'s involve on-shell only specific
combinations of the gauge parameters corresponding to the
$\7C^N$. This suggests to interpret the $\cT$'s as
tensor fields characterized through the transformation law
(\ref{e13}). Accordingly, the algebra (\ref{e10}) is
interpreted as a gauge covariant
algebra on tensor fields.

Note that the total derivative operators
$\6_\mu$ might look quite complicated when expressed
in terms of the new jet coordinates.
(\ref{e14}) shows how they are realized on the $\cT$'s, 
up to weakly vanishing terms. This realization suggests
to interpret the $\cA_\mu^N$ as connections, as
they relate partial derivatives of tensor fields
to gauge covariant quantities.
(\ref{cov15}) supports this interpretation as it
is a transformation law typical of a connection.
Moreover, very often (\ref{e14}) allows us to define covariant 
derivatives. Indeed, assume that the set $\{\cA_\mu^N\}$ 
contains an invertible subset $\{\cA_\mu^a\}$. Then one can solve 
(\ref{e14}) for the $\nabla_a$
which in turn can be interpreted as covariant derivatives as they
extend the $\6_\mu$ to covariant operations.

(\ref{cov16}) is reminiscent of a zero-curvature condition.
In fact it can be interpreted in this way as it is a consequence of 
$[\6_\mu,\6_\nu] =0$, evaluated on the $\cT^r$ by means of (\ref{e14}). 
However, usually it has a more familiar and useful interpretation,
provided $\{\cA_\mu^N\}$ contains an invertible subset $\{\cA_\mu^a\}$ 
(see above). Namely in such cases (\ref{cov16}) can be solved for 
the ${\cF_{ab}}^M$ which are then interpreted
as the curvatures corresponding to the $\nabla_a$, as one has
$[\nabla_a,\nabla_b]\approx -{\cF_{ab}}^M\nabla_M$.
\medskip

{\sc Remarks.}
a) (\ref{e2}) can fail to be valid globally if the manifold 
of the $\cU$'s has nontrivial de Rham cohomology 
within the space of local total forms; such global aspects
can be taken into account using (\ref{kunneth}), cf.\ \cite{bbh3} for an
example.

b) Often (but not always) one can choose new local jet coordinates
such that all $\cW$'s have nonnegative total degrees. 
It can also happen that
there are $\cW$'s with negative total degrees but they do not
contribute nontrivially to the cohomology. In such cases
the antifields enter the solutions of the cohomological
problem only via the antifield dependent terms
contained in $\4\cC$'s and $\4\cT$'s. The example treated
in the next section is of this type.

c) Recall that (\ref{e10})--(\ref{cov16}) arose directly
from (\ref{e1}). Conversely one may regard
a gauge covariant algebra (\ref{e10})
as the reason behind the existence of
local jet coordinates satisfying (\ref{e1}). 
Sometimes it is even fruitful to reverse the perspective
completely and to try to {\em construct}
gauge theories by imposing
an algebra (\ref{e10}) and seeking a local 
off-shell realization $\nabla_M:\cT^r\mapsto R_M^r(\cT)$.
The gauge transformations and covariant derivatives
are then obtained afterwards from (\ref{e13})--(\ref{cov15}).

\mysection{BRST cohomology for geodesic motion}
\label{example}

The approach outlined above is now illustrated for the geodesic
motion $X^m(t)$ of point particles in 
Riemannian or pseudo-Riemannian manifolds. These motions are
described by stationary points of the action functional
\beq
S=\int dt\, L,\quad L=\frac 12\left[
e^{-1}  g_{mn}(X)\, \dot X^m\dot X^n+M^2 e\right].
\label{ex1}\eeq
Here $g_{mn}$ is the metric tensor field of the (pseudo-) Riemannian
manifold, $e=e(t)$ is an auxiliary dynamical variable,
$M$ is a constant (the ``mass'' of the particle), 
and $\dot X^m$ denotes the derivative of the $X^m(t)$ with respect to 
the world-line parameter $t$. We shall assume $M\neq 0$, 
as the case $M=0$ (corresponding to null-geodesics) is slightly special.

The equations of motion are
\beq
\frac{\delta S}{\delta e}=0,\quad
\frac{\delta S}{\delta X^m}=0,
\label{eom}
\eeq
where the left hand sides are the
Euler-Lagrange derivatives of $L$
with respect to $e$ and $X^m$, 
\bea
 \frac{\delta S}{\delta e}&=&\frac 12 \left[
M^2 -e^{-2} g_{mn}(X)\, \dot X^m \dot X^n\right],
\label{ex2}\\
 \frac{\delta S}{\delta X^m}&=&
-g_{mn}(X)\left[
\frac{d}{dt}(e^{-1} \dot X^n)
+e^{-1} {\Gamma_{k\ell}}^n(X)\,\dot X^k \dot X^\ell
\right]
\label{ex3}\eea
with
\beq
{\Gamma_{mn}}^k(X)=\frac 12\, g^{k\ell}(X)
\left[
\frac{\6 g_{n\ell}(X)}{\6 X^m}
+\frac{\6 g_{m\ell}(X)}{\6 X^n}
-\frac{\6 g_{mn}(X)}{\6 X^\ell}
\right].
\label{ex4}
\eeq
The equations of motion are related by a Noether identity
equivalent to the gauge invariance of
(\ref{ex1}) associated with
reparametrizations of the world-line,
\beq
\dot X^m\, \frac{\delta S}{\delta X^m}
-e\, \frac{d}{dt} \frac{\delta S}{\delta e}=0.
\label{ex5}\eeq
As the first Eq.\ (\ref{eom}) can be solved algebraically
for $e$, one can eliminate $e$ and work from the beginning
with the $X^m$ as the only dynamical variables (the 
corresponding form of the Lagrangian is then
$M|g_{mn}(X)\, \dot X^m \dot X^n|^{1/2}$).
However, for our purposes it is more convenient
to perform first the cohomological analysis and eliminate $e$
afterwards through the substitution
\[
e \quad \rightarrow \quad 
M^{-1}\sqrt{|g_{mn}(X)\, \dot X^m\, \dot X^n|}\ .
\]

As the equations of motion derive from an action functional,
fields and antifields are in one-to-one correspondence
(``conjugate''),
\beq
\{\Phi^A\}=\{X^m\, ,\, e \, ,\, C\},\quad
\{\Phi^*_A\}=\{X^*_m\, ,\, e^* \, ,\, C^*\},
\label{ex6}\eeq
where $C$ is the ghost field associated with world-line 
reparametrizations. Thus, the standard jet coordinates
(\ref{standard}), supplemented with the
only differential $dt$, are in this case
\beq \{Z^M\}=\{t\, ,\, dt\, ,\, \6^q\Phi^A\, ,\, \6^q\Phi^*_A:\
q=0,1,\dots\},
\label{jet}\eeq
where $\6$ represents differentiation with respect to $t$.
The BRST transformations are
\bea
& & s\, t=0,\quad s\, dt=0,
\nonumber\\
& & s\, e = \6 (C e),\quad 
s\, X^m = C\6 X^m,\quad
s\, C=C\6 C,
\nonumber\\
& & s\, e^* = \frac{\delta S}{\delta e} + C\6 e^*,\quad
s\, X^*_m = \frac{\delta S}{\delta X^m} + \6(CX^*_m),
\nonumber\\
& & s\, C^* = X^*_m\6 X^m - e\6 e^* + 2C^*\6 C + C\6 C^*.
\label{sZ}
\eea
The BRST transformations of the remaining jet coordinates
(\ref{jet}) are obtained from (\ref{sZ}) simply by prolongation,
using $s\6 =\6 s$.

For definiteness, we shall now compute the BRST cohomology
in the space of functionals whose integrands are polynomials
in all jet coordinates (\ref{jet}) except for the (undifferentiated)
$X^m$ and $e$ on which they can depend arbitrarily. Accordingly
we define the space of local total forms, though the arguments
and results can be easily adapted to more general spaces of
functionals and forms. 
\medskip

{\bf Definition.} {\em The space $\4\Omega^*$ of
local total forms is defined as the space of functions $f(Z)$ depending
arbitrarily on $e$ and the $X^m$,
but polynomially on all other $Z^M$.}
\medskip

To compute $H(\4s,\4\Omega^*)$
we seek new jet coordinates satisfying (\ref{e1}).
(\ref{sZ}) suggests
\beq
\{\cU^\ell\}=\{t\, ,\, \6^q e\, ,\, \6^q X^*_m\, ,\, \6^q C^*:\
q=0,1,\dots\}
\label{Us}\eeq
because the $\4s$-variations of these $\cU$'s form a set 
$\{\cV^\ell\}$ of admissible new jet coordinates substituting
for $\{dt,\6^{q+1} C,\6^{q+2}X^m,\6^{q+1}e^*:\ q=0,1,\dots\}$, 
as one has
\beann
& \4s\, t=dt,\quad \4s\, \6^q e=e\6^{q+1} C+\dots\ ,&
\\
& \4s\, \6^q X^*_m=-e^{-1}g_{mn}(X)\6^{q+2}X^n+\dots\ ,\quad
\4s\, \6^q C^*=-e\6^{q+1}e^*+\dots\ . &
\eeann
So far we have substituted new
local jet coordinates $\{\cU^\ell,\cV^\ell\}$
for all standard jet coordinates
(\ref{jet}) except for $C$, $X^m$, $\6X^m$ and $e^*$.
In order to apply the reasoning of the previous section,
one needs in addition new jet coordinates $\cW^i$ substituting
for the $X^m$, $\6X^m$, $C$ and $e^*$ such that
(\ref{e1}) holds. This is fulfilled by
\bea
\{\cW^i\}&=&\{X^m\, ,\, P^m\, ,\, \4\cC\, ,\, \4e^*\},
\nonumber\\
\4\cC &=& e\, (C+dt),
\nonumber\\
P^m &=& e^{-1}\6 X^m - (C+dt)\, g^{mn}(X)\, X^*_n\ ,
\nonumber\\
\4e^* &=& e^* - e^{-1}(C+dt)C^* .
\label{Ws}\eea
Indeed, as required by (\ref{e1}), the $\4s$-transformations of
$X^m$, $P^m$, $\4\cC$ and $\4e^*$
can be expressed completely in terms of these variables again,
\bea
& \4s\, X^m = \4\cC P^m, \quad
\4s\, P^m = -\4\cC P^k P^\ell\,{\Gamma_{k\ell}}^m(X), &
\nonumber\\
& \4s\, \4\cC = 0,\quad
\4s\, \4e^* = \frac12\, ( M^2-\langle P \rangle^2), &
\label{sW}
\eea
where we used the notation
\[
\langle P \rangle^2 := g_{mn}(X)P^m P^n.
\]
In the terminology of section \ref{tensors},
$X^m$ and $P^m$ are generalized tensor fields $\4\cT^r$, and
$\4\cC$ is the only generalized connection.
The corresponding covariant operation (\ref{e8}) is
\beq
\4\nabla=P^m\frac{\6}{\6X^m}
-P^kP^\ell\,{\Gamma_{k\ell}}^m(X)\frac{\6}{\6P^m}\ .
\label{D}\eeq
We shell see that the kernel of $\4\nabla$ determines
constants of the motion. Note that
$\4s$ is realized on (functions of)
the new jet coordinates through
\beq
\4s=\cV^\ell\frac{\6}{\6\cU^\ell}+\4\cC\4\nabla+
\frac 12\, (M^2-\langle P \rangle^2)\frac{\6}{\6\4e^*}\ .
\label{4saction}\eeq
$\4s^2=0$ holds because of
\beq
\4\nabla \langle P \rangle^2=0.
\label{onlyid}\eeq
It is now straightforward to derive the following result.
\bigskip

{\bf Lemma.} {\em For $M\neq 0$,
$H(\4s,\4\Omega^*)$ can be represented
solely by local total forms $\4\cC G(X,P)$ and 
$F_0(X,P)\in Ker(\4\nabla)$,
\beq
M\neq 0:\ \4s\,\4\omega=0\ \Leftrightarrow\
\4\omega=\4s\,\4\eta+F_0(X,P)+\4\cC G(X,P),
\quad \4\nabla F_0(X,P)=0.
\label{res1}
\eeq
Here $F_0$ and $G$ are defined up to
the following shifts corresponding to cohomologically trivial
redefinitions of $F_0(X,P)$ and $\4\cC G(X,P)$,
\bea
F_0(X,P) & \rightarrow & F_0(X,P)
-\frac 12(M^2-\langle P \rangle^2)\7H_0(X,P),
\quad\ \4\nabla \7H_0(X,P)=0,
\label{shiftF}\\
G(X,P) & \rightarrow & G(X,P)
-\4\nabla \7F(X,P)+\frac 12(M^2-\langle P \rangle^2)\7K(X,P),
\label{shiftG}\eea
where $\7H_0$, $\7F$ and $\7K$ are polynomials in the $P^m$.
In particular, $F_0(X,P)+\4\cC G(X,P)$ is trivial
in $H(\4s,\4\Omega^*)$ if and only if $F_0$ and $G$
can be removed through shifts (\ref{shiftF}) and
(\ref{shiftG}).}
\medskip

{\bf Proof.} Due to (\ref{e2}) it is sufficient
to compute $H(\4s,\4\Omega^*_\cW)$. As $\4\cC$ and $\4e^*$ are
Grassmann odd (anticommuting), one has
\beq
\4\alpha\in \4\Omega^*_\cW\ \Leftrightarrow\ 
\4\alpha=F(X,P)+\4\cC G(X,P)+\4e^* H(X,P)+\4\cC\,\4e^*\, K(X,P).
\label{ex7}\eeq
Using (\ref{4saction}) and leaving out the arguments $(X,P)$, 
one obtains
\beq
\4s \4\alpha=
\frac 12(M^2-\langle P \rangle^2)H
+\4\cC [\4\nabla F-\frac 12(M^2-\langle P \rangle^2)K]
+\4\cC\, \4e^*\, \4\nabla H.
\label{som}\eeq
This shows
\beq
\4s\4\alpha=0\quad \Leftrightarrow\quad H=0,\quad
\4\nabla F=\frac 12(M^2-\langle P \rangle^2)K.
\label{som=0}\eeq
Next we examine how $\4\alpha$ changes if we subtract an $\4s$-exact
local total form $\4s\4\beta$, $\4\beta\in\4\Omega^*_\cW$,
\beq
\4\alpha\ \rightarrow\  \4\alpha'=\4\alpha-\4s\4\beta,\quad
\4\beta=\7F+\4\cC \7G+\4e^* \7H+\4\cC\, \4e^*\, \7K
\label{triv2}\eeq
where
$\7F=\7F(X,P)$, \dots , $\7K=\7K(X,P)$.
One finds $\4\alpha' = F'+\4\cC G'+\4\cC\4e^*K'$ where
\bea
F'&=&F-\frac 12(M^2-\langle P \rangle^2)\7H,
\label{triv3}\\
G'&=&G-\4\nabla \7F+\frac 12(M^2-\langle P \rangle^2)\7K,
\label{triv4}\\
K'&=&K-\4\nabla \7H.
\label{triv5}\eea
We show now that one can always choose $\7H$ such that $K'=0$.
To this end, we decompose $F$, $K$ and $\7H$ into parts of definite
degree $\lambda$ in the $P$'s,
\[
F=\sum_{\lambda=0}^{\lambda_F} F^{(\lambda)},\quad
K=\sum_{\lambda=0}^{\lambda_K} K^{(\lambda)},\quad
\7H=\sum_{\lambda=0}^{\lambda_{\7H}} \7H^{(\lambda)}.
\]
As $\4\nabla$ raises the degree in the $P$'s by one,
(\ref{som=0}) requires
\beq
\4\nabla F^{(\lambda)}=\frac 12\,
\left[ M^2\, K^{(\lambda+1)}-\langle P \rangle^2 K^{(\lambda-1)}\right].
\label{la}\eeq
In particular this implies $K^{(0)}=0$. Assume now
that $K^{(\lambda)}=0$ holds for all $\lambda\leq \lambda_0$.
(\ref{la}) then gives
$K^{(\lambda_0+1)}=2M^{-2}\4\nabla F^{(\lambda_0)}$.
Inserting this in (\ref{triv5}), one sees that 
$\7H^{(\lambda_0)}=2M^{-2}F^{(\lambda_0)}$ implies
$K'{}^{(\lambda_0+1)}=0$.
Since this applies to all $\lambda_0$ and is satisfied
for $\lambda_0=0$, we can indeed choose $\7H$ so as to achieve 
$K'=0$. Actually the procedure
removes simultaneously all parts of $F'$ up to and including
order $(\lambda_K-1)$.
However, in general one cannot remove $F'$ completely
because by removing $F'{}^{(\lambda)}$ one
modifies $F'{}^{(\lambda+2)}$, as is seen from (\ref{triv3}).
Hence, the procedure shifts the dependence of
$F'$ on the $P^m$ to terms of higher degree. This can be
avoided by refining the procedure.
Namely, in order to remove $K'{}^{(\lambda_0+1)}$, it
is actually sufficient to choose
\[
\7H^{(\lambda_0)}=2M^{-2}(F^{(\lambda_0)}-F_0^{(\lambda_0)}),\]
where $F_0^{(\lambda_0)}$ is the zero mode of $\4\nabla$
contained in $F^{(\lambda_0)}$.
Clearly, this zero mode does not contribute in (\ref{triv5}).
One can thus achieve $K'=0$ without modifying the zero
modes of $\4\nabla$ contained in $F$. 
Having removed $K'$, one is
left with the zero modes of $\4\nabla$ contained in $F$ and
has still the freedom to redefine the latter by choosing
$\7H\in Ker(\4\nabla)$ without violating $K'=0$.
This yields (\ref{shiftF}) which shows in particular
that $F_0$ and $\langle P \rangle^2 F_0$ are cohomologically
equivalent (choose $\7H_0=2 M^{-2} F_0$) and explains thereby
why one can shift the dependence of $F'$ on the $P^m$ to higher orders.
(\ref{shiftG}) follows directly from (\ref{triv4}).
This proves the lemma.\hspace*{\fill} $\Box$
\bigskip

Note that $F(X,P)$ and $\4\cC G(X,P)$ have total degree 0 and 1
respectively, and recall that the BRST cohomology
on local functionals at ghost number $g$ is
isomorphic to $H(\4s,\4\Omega^{g+1})$ in this case (due to $n=1$).
We conclude that the BRST cohomology on local functionals
is trivial at all ghost numbers different from
$-1$ and 0, and is represented at ghost numbers $-1$ and 0 by the
integrated 1-forms contained in $F(X,P)$ and $\4\cC G(X,P)$.
To extract these representatives, we recall that 
$P^m=\cD X^m-(C+dt) g^{nm}(X)X^*_n $ with
\beq
\cD X^m:=e^{-1}\6 X^m.
\label{DX}\eeq
As $(C+dt)$ is Grassmann odd and thus squares to zero, one has
\bea
& F_0(X,P)=F_0(X,\cD X)
-(C+dt)\, X^*_m\, F_0^m(X,\cD X), &
\label{FG1}\\
& \4\cC G(X,P) = (C+dt)\, e\, G(X,\cD X) &
\label{FG2}\eea
where
\beq
F_0^m(X,\cD X)=g^{mn}(X)\, \frac{\6 F_0(X,\cD X)}{\6\cD X^n}\ .
\label{Fm}\eeq
{}From (\ref{FG1}) and (\ref{FG2}) one reads off the BRST invariant 
functionals arising from $F_0(X,P)$ and $\4\cC G(X,P)$,
\bea
W^{-1}&=&-\int dt\, X^*_m\, F_0^m(X,\cD X),
\label{-1}\\
W^0&=&\int dt\, e\, G(X,\cD X).
\label{0}
\eea

{\bf Discussion of the result.}
According to the interpretation of the local BRST
cohomology at ghost numbers $-1$ and 0 \cite{bbh1,bh},
$W^{-1}$ and $W^0$ yield rigid symmetries and
consistent deformations of
an action functional (\ref{ex1}) respectively.

The rigid symmetries read off from $W^{-1}$ are generated by
\beq 
\delta_\epsilon e=0,\quad
\delta_\epsilon X^m=\epsilon\, F^m_0(X,\cD X)
\label{rigid} \eeq
where $\epsilon$
is a constant infinitesimal parameter. Indeed, thanks
to $\4\nabla F_0(X,P)=0$, the Lagrangian
(\ref{ex1}) transforms under (\ref{rigid}) into
a total derivative,
\beq
\4\nabla F_0(X,P)=0\quad \Rightarrow\quad 
\delta_\epsilon L = \epsilon\, \6 \left[
\cD X^m\, \frac{\6 F_0(X,\cD X)}{\6\cD X^m}-F_0(X,\cD X)
\right].
\label{symm}\eeq
The constants of the motion corresponding to these symmetries 
according to Noether's first theorem \cite{noether} are simply
the functions $F_0(X,\cD X)$ themselves. Indeed,
\beq 
\4\nabla F_0(X,P)=0\quad \Rightarrow\quad
\6 F_0(X,\cD X) \approx 0, \label{constant}\eeq
i.e.\ $F_0(X(t),e^{-1}(t)\dot X(t))$ is constant for any solution 
to the equations of motion. 
In fact, these statements apply to any zero mode of $\4\nabla$,
whether or not it depends polynomially on the $P^m$ 
(since $\4\nabla F_0(X,P)=0 \Leftrightarrow \4s F_0(X,P)=0$).

Hence, each zero mode of $\4\nabla$ provides a 
rigid symmetry of (\ref{ex1}) and a corresponding constant
of the geodesic motion. However, not all these
symmetries and constants of the motion are independent and some
are trivial. In fact,
two such symmetries or constants of the motion are equivalent if
they are related through a redefinition (\ref{shiftF}). In particular,
symmetries and constants of the motion are trivial when
$F_0(X,P)=\frac 12(M^2-\langle P \rangle^2)\7H_0(X,P)$ for some 
$\7H_0(X,P)\in Ker(\4\nabla)$ because then
(\ref{rigid}) reduces on shell to a world-line reparametrization
and the associated constant of the motion just vanishes on-shell
due to $M^2-\langle \cD X \rangle^2\approx 0$.
For instance, $F_0=M^2-\langle P \rangle^2$
results in $\delta_\epsilon X^m=-2\epsilon e^{-1}\6 X^m$
which represents an infinitesimal world-line reparametrization
$t\rightarrow t-2\epsilon\, e^{-1}(t)$.

Those $F_0(X,P)$ which depend polynomially on the $P^m$ correspond
to the Killing tensors of the Riemannian manifold.
Indeed, consider
\beq F_0^{(\lambda)}(X,P) = P^{m_1}\dots P^{m_\lambda}
f_{m_1\dots m_\lambda}(X).
\eeq
It is easy to verify that $\4\nabla F_0^{(\lambda)}(X,P)=0$
is equivalent to
\beq
D_{(m_0}\, f_{m_1\dots m_\lambda)}(X)=0,
\label{covconst}\eeq
where $D_m$ is the usual covariant derivative  
with connection ${\Gamma_{mn}}^k$ and
$(m_0\dots m_\lambda)$ denotes complete
symmetrization. The solutions of (\ref{covconst})
are called Killing tensors of order $\lambda$ since
(\ref{covconst}) generalizes the Killing vector equations
(recovered for $\lambda=1$). We have thus reproduced the
well-known result that the Killing tensors give rise to
constants of geodesic motion, see e.g.\ \cite{kramer}.

Let us finally discuss the deformations (\ref{0}).
Note that they contain only first order derivatives.
This does not mean that there are no reparametrization invariant
local functionals containing higher order derivatives; in fact
one can easily construct such functionals (see below).
Rather, the cohomology tells us that any such functional is
cohomologically equivalent to one of the form (\ref{0}),
i.e.\ it can be brought to this form 
by subtracting BRST exact local functionals.
This implies that a reparametrization invariant
deformation of the action functional (\ref{ex1}) with
deformation parameter $\tau$,
\[ S_\tau = S+\tau S_1+O(\tau^2)\ , \]
can always be
rewritten such that $S_1$ takes the form (\ref{0})
by means of local redefinitions of the dynamical variables.
For instance, the functional
\[
S_1=\int dt\, e\, h_{mn}(X)\cD^2 X^m\, \cD^2 X^n,\quad
\cD^2 X^m=e^{-1}\6(e^{-1} \6 X^m)
\]
is reparametrization invariant. It is indeed cohomologically
equivalent to a functional $S'_1$ of the form (\ref{0}),
\beann
S_1 &=& S'_1 + s \int dt\, X^*_m\, g^{mn}(X)\, h_{nk}(X)\, 
[{\Gamma_{rs}}^k(X)\cD X^r\,\cD X^s-\cD^2 X^k],
\\
S'_1 &=& \int dt\, e\, 
h_{mn}(X)\, {\Gamma_{k_1 k_2}}^m(X)\, {\Gamma_{k_3 k_4}}^n(X)
\cD X^{k_1}\dots  \cD X^{k_4}
\eeann
where we neglected total derivatives in the integrand.
Accordingly one has
\[
S[X,e] + \tau S_1[X,e] = S[X',e] + \tau S'_1[X',e] + O(\tau^2)
\]
where
\[
X'{}^m = X^m + \tau g^{mn}(X)\, h_{nk}(X)\, 
[{\Gamma_{rs}}^k(X)\cD X^r\,\cD X^s-\cD^2 X^k].
\]

{\sc Remarks.}
a) $\4\nabla$ can be used to define a covariant derivative $\nabla$
on tensor fields $\cT^r$, as explained at the end of 
section \ref{tensors}. The tensor fields are
the antifield independent parts of the $\4\cT^r$,
i.e.\ $\{\cT^r\}=\{X^m\, ,\, \cD X^m\}$.
The covariant derivative acts as
\[
\nabla X^m=\cD X^m,\quad
\nabla^2X^m=-{\Gamma_{k\ell}}^m(X)\,\nabla X^k\, \nabla X^\ell,
\quad etc.
\]
Note that $\nabla^q X^m$ has degree $q$ in the $\cD X^m$.

b) The general form of reparametrization invariant local
functionals containing higher order derivatives of the $X^m$ is
\beq \int dt\, e\, B(X,\cD X,\cD^2X,\dots,\cD^q X),\quad
\cD=e^{-1}\6
\label{BB}\eeq
where $B$ is an arbitrary function of its arguments.
This can be proved by computing analogously the cohomology of $\4s$ 
in the space of antifield independent local total forms.
In this space the set of tensor fields contains all
the $\cD^q X^m$ ($q=0,1,\dots$) and the covariant derivative 
is $\cD=e^{-1}\6 $. The cohomological equivalence
of a functional (\ref{BB}) to one of the form (\ref{0}) follows
from the fact that $\cD$ and $\nabla$ coincide on-shell,
\[ \cD X^m=\nabla X^m,\quad
\cD^q X^m\approx \nabla^q X^m\quad (q=2,3,\dots). \]

{\sc Acknowledgement.} The author was supported by the
Spanish ministry of education and science (MEC).

\end{document}